\documentclass[aps,prb,preprint,superscriptaddress,endfloats*,showpacs,showkeys]{revtex4-1}

\usepackage{graphicx}
\usepackage{dcolumn}
\usepackage{bm}
\newcommand{\vecvar}[1]{\mbox{\boldmath$#1$}}

\begin{document}

\preprint{PRESAT-8402}

\title{First-principles electronic-structure calculation of dangling bonds at Si/SiO$_2$ and Ge/GeO$_2$ interfaces}

\author{Tomoya Ono}
\affiliation{Department of Precision Science and Technology, Osaka University, Suita, Osaka 565-0871, Japan}
\affiliation{Research Center for Ultra-Precision Science and Technology, Osaka University, Suita, Osaka 565-0871, Japan} 

\author{Shoichiro Saito}
\affiliation{Department of Precision Science and Technology, Osaka University, Suita, Osaka 565-0871, Japan}

\date{\today}

\begin{abstract}
Evidence of the absence of the clear electron spin-resonance signal from Ge dangling bonds (DBs) at Ge/GeO$_2$ interfaces is explored by means of first-principles electronic-structure calculations. Comparing the electronic structures of the DBs at Si/SiO$_2$ and Ge/GeO$_2$ interfaces, we found that the electronic structure of the Ge-DB is markedly different from that of the Si-DB; the Ge-DB states does not position in the energy band gap of the Ge/GeO$_2$ interface while the Si-DB states clearly appears. In addition, the charge density distribution of the Ge-DB state spreads more widely than that of the Si-DB state. These features are explained by considering the metallic properties of the bonding network of the Ge/GeO$_2$ interface and the structural deformation of the Ge bulk at the Ge/GeO$_2$ interface due to the lattice-constant mismatch.
\end{abstract}

\maketitle

\section{Introduction}
The length scale of Si-based electronic devices has been continuously decreasing and is approaching its technological and physical limits. Thus, a considerable number of studies have been conducted to find alternative materials to further improve the performance of metal-oxide-semiconductor field-effect transistors (MOSFETs). Among the possible candidate materials, Ge has attracted considerable attention owing to Ge-MOSFETs exhibiting higher carrier mobility than Si-MOSFETs. One of the most important issues in the application of Ge-MOSFETs is the accurate control of the properties of the Ge/GeO$_2$ interface because it exists even at a Ge/high-{\it k} oxide interface. Electron spin-resonance (ESR) investigations have revealed that interfacial dangling bond (DB) defects at Ge/GeO$_2$ interface play a different role from those at Si/SiO$_2$ interfaces; \cite{afanasev} a measurable density of DBs of the semiconductor surface atoms is not found at Ge/GeO$_2$ interfaces although Si-DBs are ESR-active.

Very recently, Matsubara {\it et al.} \cite{matsubara} and Hosoi {\it et al.} \cite{hosoi} demonstrated that the interface defect density at Ge/GeO$_2$ interfaces is lower than that at Si/SiO$_2$ interfaces when both the interfaces are fabricated with no hydrogen passivation treatment. Houssa {\it et al.} \cite{houssa} examined by first-principles calculation the density of DBs at the Ge/GeO$_2$ interface as a function of oxidation temperature by combining viscoelastic data for GeO$_2$ and the modified Maxwell's model, and they addressed the lower defect density at Ge/GeO$_2$ interfaces than at Si/SiO$_2$ interfaces. In addition, the present authors performed a first-principles calculation of the formation energy of defects at the Ge/GeO$_2$ interface during oxidation and found that hardly any defects are generated at the interface compared with the number generated at the Si/SiO$_2$ interface.\cite{saito} These theoretical calculations are in agreement with the above experiments,\cite{matsubara,hosoi}and the low defect density at the interface may account for the absence of DBs observed in ESR experiments, consistent with the conclusions drawn in these studies.

On the other hand, from the viewpoint of the electronic structure of DBs, Weber {\it et al.} \cite{weber} investigated the electronic structure of DBs in Ge bulk by first-principles calculations and found that Ge-DBs have an energy level below the Ge valence band edge; Ge-DBs are always negatively charged and fully occupied. Broqvist {\it et al.} \cite{broqvist} also examined the charged states of the DBs in Ge bulk and found two charge-transition levels in the lower part of the band gap, indicating the occurrence of three charged states in analogy with the DBs in Si, using the hybrid functional, although these two levels are below the valence band edge in the case of the local density approximation. However, Weber {\it et al.} \cite{weber} have also proved that the underestimation of the energy band gap due to the local density approximation does not cause the negatively charged state of DBs by employing the $G_0W_0$ approximation. Houssa {\it et al.} \cite{houssa} reported that the defect state associated with the DBs lies approximately in the middle of the Ge energy band gap using a model in which GeO$_2$ is piled on a Ge(001) substrate. However, they employed pseudopotentials including a repulsive potential so that the energy band gap of Ge agrees with the experimental value, and it is unclear the extent to which the deformation of the pseudopotentials affects the atomic and electronic structures at the Ge/GeO$_2$ interface. Although it is difficult to adhere to one of these views solely owing to the use of approximations in these theoretical calculations, the examination of the difference between the electronic structures of the DBs at Si/SiO$_2$ and Ge/GeO$_2$ interfaces is of importance for clarifying the electronic structure of DBs and for the realization of Ge-based devices.

In this study, we investigate the electronic structures of DBs at the (001)Ge/GeO$_2$ interface compared with those at the (001)Si/SiO$_2$ interface by first-principles calculations to determine the reason for the absence of the distinct ESR signal from DBs at the Ge/GeO$_2$ interface. Our findings are as follows. The electronic structure of the Ge-DB is notably different from that of the Si-DB. The Ge-DB state is occupied at any position of the Fermi level, whereas the Si-DB is paramagnetic when the Fermi level is near the midgap. In addition, the Ge-DB state couples with the interface state and its charge density distribution extends more widely than that of the Si-DB state.

\section{Computational Methods}
\begin{figure}
\includegraphics{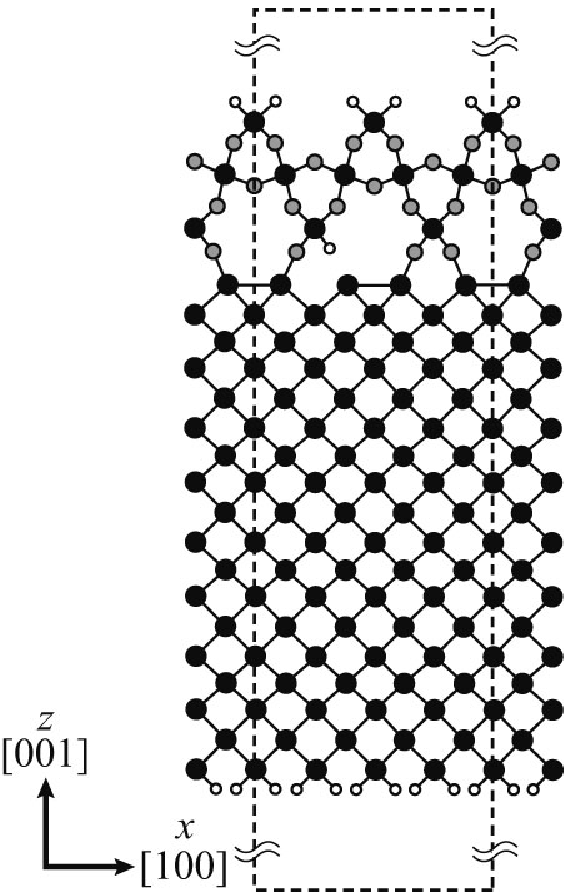}
\caption{\label{fig:fig1} Schematic image of computational model. Black, gray, and white circles represent Si (Ge), O, and H atoms, respectively. The rectangle enclosed by broken lines represents the supercell.}
\end{figure}

Our first-principles calculation method is based on the real-space finite-difference approach,\cite{rsfd,icp,tsdg} which enables us to determine a self-consistent electronic ground state with a high degree of accuracy using a timesaving double-grid technique.\cite{icp,tsdg} Intensive studies by various methods have shown that the crystalline phase of SiO$_2$ can be observed down to $\sim$ 10 \AA \hspace{1mm} from the interface in some cases.\cite{intefaceexp1,intefaceexp2,intefaceexp3,intefaceexp4} Although no crystalline phase of GeO$_2$ has yet been experimentally observed, we adopt the same atomic configuration for the Ge/GeO$_2$ interface to compare the electronic structure of DBs with that of DBs at the Si/SiO$_2$ interface. Our interface structure corresponds to that proposed by Kageshima and Shiraishi for the (001)Si/SiO$_2$ interface during oxidation.\cite{model} Figure~\ref{fig:fig1} shows the computational model, consisting of an $\alpha$-quartz crystalline oxide layer on a (001) substrate of $\sim$23 \AA\hspace{1mm} thickness. Since Yamasaki {\it et al.}\cite{yamasaki} claimed that an effective band-gap change from Si to SiO$_2$ occurs on the oxide side within 5 \AA\hspace{1mm} from the Si/SiO$_2$ interface, the thickness of the oxide layer is chosen to be $\sim$ 8 \AA. The size of the supercell is taken to be 2$a_0 \times$ 2$a_0 \times$ 5$a_0$, where $a_0$ is the experimental lattice constant of Si (5.43 \AA) or Ge (5.65 \AA). The periodic boundary condition is imposed in the $x$ and $y$ directions, while the isolated boundary condition is employed in the $z$ direction. The DBs at the interface are generated using the procedure employed by Houssa {\it et al.};\cite{houssa} a bridging O atom is removed and the lower DB is passivated with a H atom. The norm-conserving pseudopotentials \cite{norm} of Troullier and Martins \cite{tmpp} are used to describe the electron-ion interaction and are transformed into the computationally efficient Kleinman-Bylander separable form,\cite{kleinman} using the $s$ and $p$ components as nonlocal components for O and Si and the $s$, $p$, and $d$ components as nonlocal components for Ge. Exchange correlation effects are treated by a local spin density approximation,\cite{lda} and the freedom of spin is considered. The cutoff energy is set to 112 Ry, which corresponds to a grid spacing $h$ of 0.30 a.u., and a denser grid spacing of $h/3$ is taken in the vicinity of the nuclei with the augmentation of double-grid points.\cite{tsdg,icp} Eight-$k$ points in the $1 \times 1$ lateral unit cell are used for Brillouin zone sampling. Structural optimization is implemented for all atoms except the Si (Ge) atoms in the bottom layer and the H atoms on the Si (Ge) surface side, reaching a tolerance of the force of $F_{max} < 0.1$ eV/\AA.

\begin{figure*}
\includegraphics{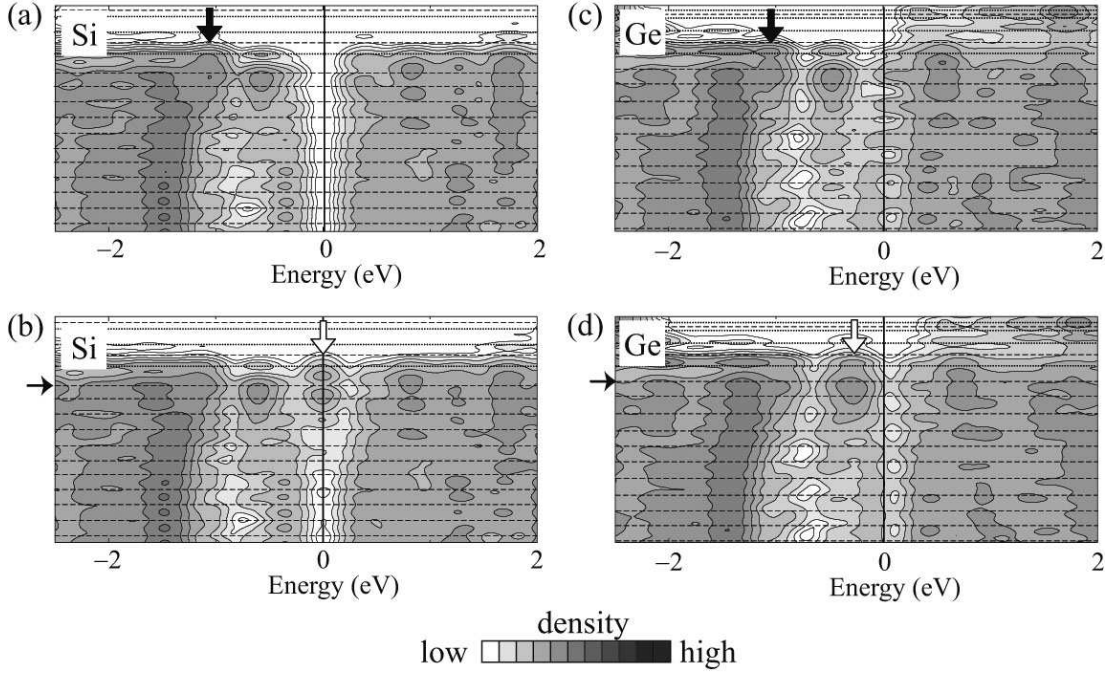}
\caption{\label{fig:fig2} Distributions of DOS integrated on plane parallel to the interface as functions of relative energy from the Fermi level. (a) Si/SiO$_2$ without DBs, (b) Si/SiO$_2$ with DBs, (c) Ge/GeO$_2$ without DBs and (d) Ge/GeO$_2$ with DBs. Zero energy is chosen to be the Fermi level. Each contour represents twice or half the density of the adjacent contour lines, and the lowest contour is 1.45 $\times 10^{-4}$ {\it e}/eV/\AA. The dashed and dotted lines represent the vertical positions of Si (Ge) and O atomic layers, respectively. The right arrows denote the position of the Si and Ge atoms forming DBs. Gaussian broadening of 0.22 eV is used and the band gap of Ge is smeared out by the broadening.}
\end{figure*}

\begin{figure}
\begin{center}
\includegraphics{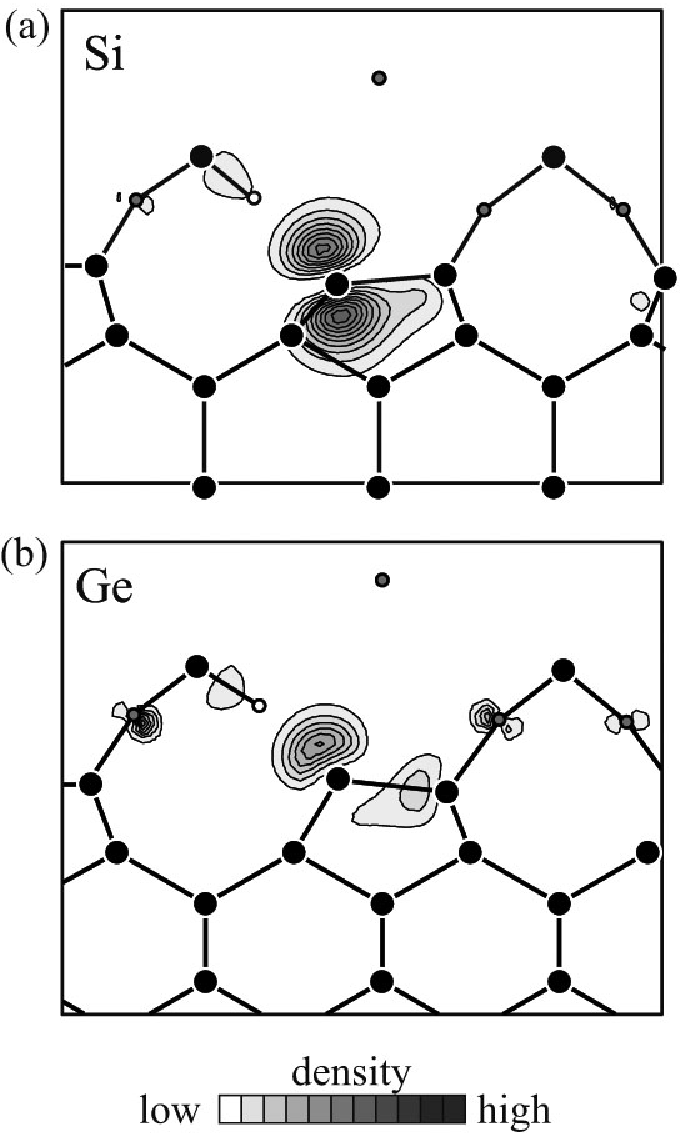}
\end{center}
\caption{ Contour plots of charge density distributions of DB states. The planes shown are along the cross section in the (110) plane including DBs. (a) Si-DB state at $E_F$ and (b) Ge-DB state at $E_F-0.3$ eV. The lowest contours are $6.75\times 10^{-3}$ $e$/\AA$^3$ and the subsequent contour lines represent values larger than $6.75\times 10^{-3}$ $e$/\AA$^3$. The meanings of the symbols are the same as those in Fig.~\ref{fig:fig1}.}
\label{fig:fig3}
\end{figure} 

\section{Results and Discussion}
Figure~\ref{fig:fig2} shows the densities of states (DOS) of the interfaces without the DB, which are plotted by integrating them along the plane parallel to the interface, $\rho(z,E)=\int |\psi(\mbox{\boldmath$r$},E)|^2 d\mbox{\boldmath$r$}_{||}$, where $\vecvar{r}=(x,y,z)$, $\psi$s are the wave functions, and $E$ is the energy of the states. For comparison, the DOS of the interfaces without any DBs is also depicted. Since we employ a model with a finite-thickness Si (Ge) substrate, the energy band gap is overestimated owing to the confinement effect; the energy band gaps of Ge/GeO$_2$ and Si/SiO$_2$ are $\sim$0.20 eV and $\sim$0.62 eV, respectively. However, Gaussian broadening of 0.22 eV is used to plot Fig.~\ref{fig:fig2} and the band gap of Ge is smeared out by the broadening in Fig.~\ref{fig:fig2}. In both the Si/SiO$_2$ and Ge/GeO$_2$ interfaces, the DOS deep inside the substrates is not affected by the presence of the DBs. In addition, the peaks of the DOS are observed at $E_F-0.3$ eV, $E_F-1.0$ eV, $E_F-1.5$ eV, and $E_F-2.0$ eV, where $E_F$ is the Fermi level. The states at $E_F-1.0$ eV indicated by the black down arrows are related to the Ge- (Si-) O bonds since their charge density distributions accumulate between the Ge- (Si-) and O-atom layers, and the bunches at $E_F-0.3$ eV and accumulating just below the interfaces are the contribution of the interface states. When a DB is introduced in the Si/SiO$_2$ interface, the DOS at $E_F-1.0$ eV decreases. Instead, a state appears in the energy band gap, indicated by the white down arrow, and the Fermi level is positioned at the state. The midgap state is attributed to the $\pi$ electron of the Si atom because its charge density distribution exhibits bunches above and below the Si atom. On the other hand, the peak of the DOS at $E_F-1.0$ eV related to the Ge-O bonds becomes shallow and the DOS at $E_F-0.3$ eV, denoted by the white down arrow, increases in the case of the Ge/GeO$_2$ interface. In addition, the Fermi level is at the valence band edge of the Ge substrate. This result implies that the Ge-DB state of the Ge/GeO$_2$ interface does not have unpaired electrons when the Fermi level is in the energy band gap of the Ge substrate, which agrees well with the conclusion of another first-principles calculation indicated that the Ge-DB state in Ge bulk is below the valence band edge of Ge bulk.\cite{weber} The charge neutrality of the Ge substrate might affect the Ge-DB state because one electron is removed from the Ge substrate by the Ge-DB. However, no significant differences were found in the DOS when additional Ge layers are attached and when one electron is added to the present system.

We illustrate in Fig.~\ref{fig:fig3} the charge density distributions of the state at the Fermi level of the Si/SiO$_2$ interface and the state below 0.3 eV the Fermi level of the Ge/GeO$_2$ interface, which correspond to the DB states corresponding to the DOS depicted in Fig.~\ref{fig:fig2}. In the case of the Si/SiO$_2$ interface, the charge density distribution of the $P_b$ center \cite{pbcenter} is in agreement with the results of a previous first-principles calculation.\cite{kato} The charge density distribution of the Ge-DB state is spatially more extended in the lateral direction than that of the Si-DB state, and we cannot observe the strong contribution of the $\pi$ electron of the Ge atom, which can be found in the Si-DB state. Thus, the Ge-DB state couples with the interface states of the Ge substrate. Ge is a group IV element between Si and Sn, the latter being metallic, and the rutile phase is energetically more stable than the other possible phases of GeO$_2$ and SnO$_2$\cite{rutile} whereas SiO$_2$ preferentially adopts the quartz structure, maintaining its $sp_3$ bonding networks. In addition, the Ge-O bonds at the interface do not have a strongly preferential bonding direction.\cite{saito} Thus, the bonding network of Ge/GeO$_2$ is easily deformed from the $sp_3$ structure and exhibits more metallic properties than that of Si/SiO$_2$. These characteristics of Ge contribute to the smaller energy variation of the Ge-O bond state after the breaking of the bond than that of the Si-O bond state and the coupling of the Ge-DB state to the Ge/GeO$_2$ interface states. These results are in agreement with the experimental result by the ESR microscopy,\cite{afanasev} and the absence of the clear ESR signal from Ge-DBs is attributed to the significant difference between the electronic structures of Si- and Ge-DBs. Our results also indicate that a DB near the interface generates a fixed negative charge, creating serious problems for devices that rely on the formation of an electron channel.

There remains one concern that our main conclusion might be affected by the computational models naively. For example, only a limited number of cases have been examined in the present work and the Ge-DB states will appear between the valence and conduction band edges in some interface atomic structures. Actually, it is difficult to deny the presence of the ESR-active Ge-DBs only by the present result. However, the charge density distribution of the Ge-DB state is not strongly concentrated around the Ge-DB as shown in Fig.~\ref{fig:fig3} and the ESR signal from Ge-DBs is expected to be shallow.

For the further interpretation of the electronic-structure of the DB states, let us discuss about the relationship between the DB states and the deformation of the atomic structure at the interface. The interface reconstruction is caused by the lattice-constant mismatch between Ge (Si) and GeO$_2$ (SiO$_2$). The interface Ge atom with the DB is raised up while the Si atom is pulled down so that the $\pi$ electron forms a $P_b$ center in the Si substrate. The average variations of the three bond angles formed by the interface atom and two atoms at the second interface layer from the ideal diamond structure (109.5 degree) is $+5.4$ and $-8.7$ degrees in the case of the Si/SiO$_2$ and Ge/GeO$_2$ interfaces, respectively. The large variation of the bond angles of the Ge/GeO$_2$ interfaces is interpreted by the structural properties of GeO$_2$ and SiO$_2$ under the pressure: O-Ge-O angle predominantly varies when the GeO$_2$ is compressed while Si-O-Si angle is deformed in the case of SiO$_2$.\cite{saito-jap} Moreover, Haneman reported in the study concerning the reconstruction of the Si surface that the $sp_3$ DB bonds is stabilized due to the strong contribution of the $s$ electron when the surface atom is raised up.\cite{haneman} This result also supports our conclusion for the stabilized Ge-DB state and the small contribution of the $\pi$ electrons to the Ge-DB state as shown in Fig.~\ref{fig:fig3}.

\section{Summary}
We have carried out first-principles calculations to examine the electronic structures of the DBs at (001)Si/SiO$_2$ and (001)Ge/GeO$_2$ interfaces. We found that the electronic structures of the Si- and Ge-DBs are markedly different; the Si-DB state appears at the energy band gap of the Si substrate and the position of the Fermi level corresponds to the DB state, whereas the Ge-DB state is fully occupied. Moreover, the charge density of the Ge-DB state is more widely distributed than that of the Si-DB state. These characteristics are explained by the metallic properties of the bonding network of the Ge/GeO$_2$ interface and the structural deformation of the Ge bulk at the Ge/GeO$_2$ interface. Our results are in good agreement with the finding in ESR measurement at the Ge/GeO$_2$ interface.\cite{afanasev} Since we studied only limited models of interfaces, not all the details of the actual process are included. Indeed, our results do not conteract the presence of the ESR-active Ge-DB at the Ge/GeO$_2$ interface, which have been reported by means of electrically detected magnetic resonance spectroscopy.\cite{baldovino} These findings further motivate the need for a theoretical knowledge for the relationship between the atomic structural property of Ge/GeO$_2$ intarface and the ESR activity of the Ge-DBs.

\section*{Acknowledgment}
The author would like to thank Professor Kenji Shiraishi of University of Tsukuba and Professor Kikuji Hirose, Professor Heiji Watanabe, and Professor Yoshitada Morikawa of Osaka University for reading the entire text in its original form and fruitful discussion. This research was partially supported by Strategic Japanese-German Cooperative Program from Japan Science and Technology Agency and Deutsche Forschungsgemeinschaft, by a Grant-in-Aid for Young Scientists (B) (Grant No. 20710078), and also by a Grant-in-Aid for the Global COE ``Center of Excellence for Atomically Controlled Fabrication Technology'' from the Ministry of Education, Culture, Sports, Science and Technology, Japan. The numerical calculation was carried out using the computer facilities of the Institute for Solid State Physics at the University of Tokyo, Center for Computational Sciences at University of Tsukuba, the Research Center for Computational Science at the National Institute of Natural Science, and the Information Synergy Center at Tohoku University.

\end{document}